\documentstyle[11pt,newpasp,twoside,psfig]{article}
\index{summary}
\index{instructions}
\index{template}

\def\edcomment#1{\iffalse\marginpar{\raggedright\sl#1\/}\else\relax\fi}
\reversemarginpar

\begin{document}
\title{Chandra Spectroscopy of Pulsars and their Wind Nebulae}

\author{E. V. Gotthelf \& C. M. Olbert}
\affil{Columbia Astrophysics Laboratory, 550 West 120th St, New York, NY 10027, USA}

\begin{abstract}
We present preliminary results from a systematic spectral study of
pulsars and their wind nebulae using the {\it Chandra X-Ray
Observatory.}  The superb spatial resolution of {\it Chandra} allows
us to differentiate the compact object's spectrum from that of its
surrounding nebulae.  Specifically, for six Crab-like pulsars, we
compare spectral fits of the averaged pulsar wind nebulae (PWN)
emission to that of the central core using an absorbed power-law
model. These results suggest an empirical relationship between the
bulk averaged photon indicies for the PWNe and the pulsar cores;
$\Gamma_{PWN} = 0.8 \times \Gamma_{Core} + 0.8$.  The photon indices
of PWNe are found to fall in the range of $1.3 < \Gamma_{PWN} < 2.3$.
We propose that the morphological and spectral characteristics of the
pulsars observed herein seem to indicate consistent emission
mechanisms common to all young pulsars.  We point out that the
previous spectral results obtained for most X-ray pulsars are likely
contaminated by PWN emission.
\end{abstract}

\section{Introduction}

Recent observation of pulsars associated with supernova remnants obtained 
with the {\it Chandra} X-ray observatory (Weisskopf, O'Dell \& van
Speybroeck 1996) are providing an unprecedentedly detailed view of
pulsar wind nebulae.  For the first time, emission features involving
wisps, co-aligned toroidal structures, and axial jets are fully
resolved in X-rays on arcsecond scales.  These features, similar to
those seen from the optical Crab nebula, are now found to be common to
young, energetic pulsar in supernova remnants (Gotthelf 2001).  Herein
we present preliminary spectral analysis of several PWNe observed with
{\it Chandra}, which, collectively, suggest a fundamental observational
relationship between the spectral characteristics of pulsars and their
pulsar wind nebulae.

Table 1 presents spectral results from a sample of {\it Chandra}
pulsars which have characteristic wind nebulae.  A summary of these
objects along with references and images can be found in Gotthelf
(2001). All observations were obtained with the ACIS-CCD camera which
is sensitive to X-rays in the 0.2--10~keV band with an energy
resolution of $\Delta E / E \sim 0.1$ at $1$ keV.  The on-axis point
spread function is slightly undersampled by the CCD pixels
($0.5^{\prime\prime}$) allowing us to isolate the pulsar emission from
that of the nebula. Except for N157B, which serendipitously fell on
ACIS-I0, all data were obtained with ACIS-S3.  The data were collected
in nominal spectral (``FAINT'') and timing (3.24~s) mode and reduced
and analyzed using the latest version of CIAO (CIAO 2.2/CALDB
v.2.9).  Starting with the Level 1 processed event files we corrected
the event data for CTI effects (Townsley et al. 2001) and applied the
standard Level 2 filtering criteria, then further rejected time
intervals of anomalous background rates.

\begin{table}
  \caption{Spectral Properties of Pulsars and their Wind Nebulae$^{a}$ }
  \begin{tabular}{lccccc} \hline
Remnant$^{b}$ & $\Gamma_{\rm PWN}^{\rm Averaged}$ & $\Gamma_{\rm Core}^{c}$ & $\Gamma_{\rm Pulsed}^{d}$  &  log L$_{\rm x\,NS}$ & log L$_{\rm x\,PWN}$ \\
\\ \tableline
G11.2$-$0.3       & 1.28$\pm$0.15 & 0.63$\pm$0.12 & 0.60$\pm$0.60  & 33.9 & 34.2 \\
Vela~XYZ          & 1.50$\pm$0.04 & 0.95$\pm$0.24 & 0.93$\pm$0.26  & 31.2 & 32.6 \\
Kes~75            & 1.88$\pm$0.04 & 1.13$\pm$0.11 & 1.10$\pm$0.30  & 35.2 & 36.0 \\
3C\,58            & 1.92$\pm$0.11 & 1.73$\pm$0.15 & \dots          & 33.0 &  \dots    \\
Crab~Nebula       & 2.11$\pm$0.05 & 1.63$\pm$0.09 & 1.86$\pm$0.07       & 35.9 & 37.3  \\
N157B~Nebula      & 2.28$\pm$0.12 & 2.07$\pm$0.21 & 1.60$\pm$0.35  & 35.9 & 36.1 \\
\tableline
\end{tabular}
\end{table}
\noindent{$^a$}{\footnotesize Ranked by increasing {\bf averaged} PWN photon index.} \\
\noindent{$^b$}{\footnotesize Values for the following objects were taken from the literature: Crab: Willingale et al.~(2001); N157B: Wang \& Gotthelf 1998.} \\
\noindent{$^c$}{\footnotesize The value for the Crab pulsar has been obtained from the literature. The measured photon-indices have been corrected for pile-up as discussed in the text.} \\
\noindent{$^d$}{\footnotesize Pulsed PI value references: Vela: Strickman, Harding \& Jager (1999);G11.2-0.3: Torii et al.~(1997); Kes~75: Gotthelf et al.~(2000); Crab: Pravdo, Angelini \& Harding (1997); N157B: Marshall et al.~(1998).} \\

\bigskip

For each object listed in Table 1, we extracted spectra from the
PWN, pulsar core, and background, when available, or obtained spectral
parameters from the literature as indicated.  For the core spectra,
the brightest central pixels were extracted based upon data above
4.0\,keV, where the core emission is substantially greater than that
of the surrounding nebula.  The nebula itself was then extracted
from the region above which the background was constant, excluding
the core.  Large variations in background rates and size of both the
nebula and the central source amongst observations precluded the use
of a standard extraction aperture.  Custom spectral response matrix
functions (RMFs) were provided with the CTI correction software, and 
ancillary response functions (ARFs) were created according to standard 
CIAO 2.2 procedures, using the QEU calibration files similarly provided.  
All spectra were grouped to a minimum of 50 counts per spectral bin.

We fit the resultant background-subtracted pulsar and PWN spectra with
a power-law model above $2$~keV using the latest version of {\tt
XSPEC} (v11.1).  The spectral fits to the core included a convolution
model to account for pileup effects on the spectra.  Table 1 lists the
spectral parameters obtained for each object; the values for the Crab
and N157B have been obtained from the literature (Willingale et
al.~(2001), Wang \& Gotthelf 1998, respectively). No measurement of
the pulsed spectrum for 3C58 is currently available.  Absorption
values were obtained by fitting the nebula spectra with an absorbed
power-law above 0.6\,keV, and were subsequently frozen for all future
fits.

The best-fit linear relationship between core and PWN photon indices
is shown in Figure 1.
We expect that the linear relationship seem in Fig. 1 may steepen
slightly as we more fully account for pileup effects in the spectra.
We note that the pileup corrected photon indices are likely subject
to change due to the preliminary nature of the pileup model
implementation, and suggest caution until these issues are resolved,
but expect the basic result to remain unchanged.

We also note that due to extreme pileup effects and the brightness of
its nebulae compared to the brightness of its core, SNR~0540-69 has not
been included pending a more detailed analysis of the available data.

As a check to our final spectral results and to look for any
systematic effect due to pile-up, we compare our measured power-law
indices with those obtained from the literature for the pulsed
emission for each source (Figure 2). Although they need not be the
same, it is reassuring that they agree to within measurement
errors.

\begin{figure}
\small
      \begin{minipage}[t]{0.49\linewidth}
        \psfig{file=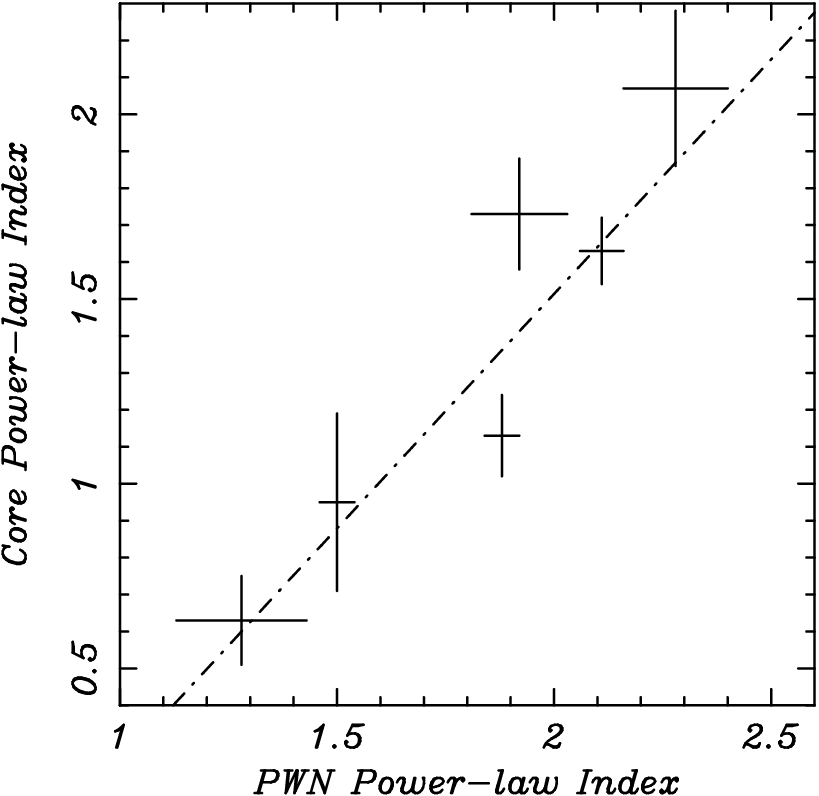,width=\linewidth,angle=270.0}
\bigskip
{
{\bf Figure 1. $\Gamma_{\rm PWN}^{\rm Averaged}$\,vs.\,$\Gamma_{\rm
Core}$} -- A plot of the power-law photon indices of the average
pulsar wind nebulae spectra versus the pileup-corrected core photon 
indices for the collection of objects presented in Table 1. A dashed-line 
indicates the best-fit linear regression (i.e.
$\Gamma_{PWN} = 0.8 \times \Gamma_{Core} + 0.8$).  The physical origin of this 
relationship is yet to be determined.
}
      \end{minipage}\hfill
      \begin{minipage}[t]{0.49\linewidth}
          \psfig{file=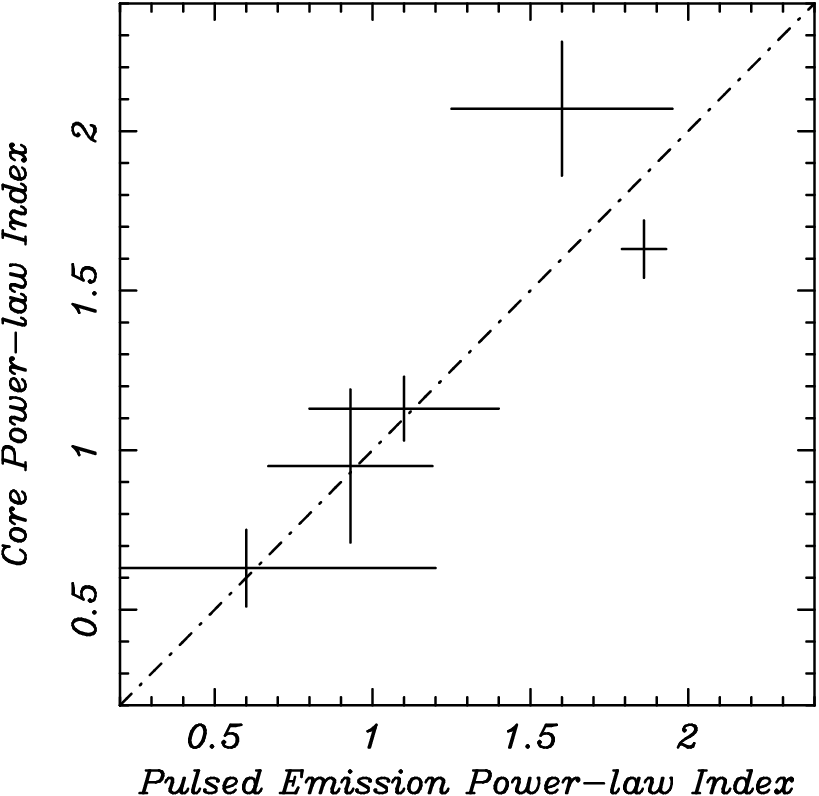,width=\linewidth,angle=270.0,clip=}
\bigskip
{
{\bf Figure 2. $\Gamma_{\rm Core}^{\rm Pulsed}$\,vs.\,$\Gamma_{\rm
Core}$} -- A comparison between the subset of measured power-law
photon index for the core emission of objects presented in Table 1 and
the pulsed emission for each of these source obtained from the
literature (from ASCA or XTE fits). A one-to-one correspondence is
indicated by the dashed line.  
}
      \end{minipage}
\end{figure}

\section{Results}

      The photon indices obtained from this analysis show a trend
that the average PWNe photon indices are consistently steeper than
their core's photon indices by a factor of one-half to one.
We suggest that this relationship is linear, with the form
$\Gamma_{PWN} = 0.8 \times \Gamma_{Core} + 0.8$.
This trend may be due to a number of factors.  It may imply that the
synchrotron spectrum frequency break occurs close to the neutron 
star, that there is a fundamental difference in emission mechanisms 
between the two regions, or that the nebula is being powered by the 
core and that the steepening is due to synchrotron burnoff or a 
comparable aging mechanism.

      {\it Chandra} is the first telescope able to spatially resolve
the neutron star from the nebula.  Previous spectral observation of
pulsars are likely to be strongly contaminated by PWN emission.  For
example, we see no clear relationship between the power-law spectral
index obtained for these pulsars by {\it ASCA} with any of those measured
by {\it Chandra}.  As more crab-like pulsars are observed and analyzed,
a more complete picture of their morphological and spectral 
characteristics comes into view, paving the way for future models and
simulations.









\acknowledgments

{This work made possible by NASA LTSA grant NAG~5-7935.  C.M.O. also acknowledges
 the support of the I.I. Rabi Scholars Program at Columbia University.}

\end{document}